\documentclass[aps,12pt,superscriptaddress,amsfonts,amssymb,amsmath]{revtex4}

\usepackage{graphicx}
\usepackage{epsfig}
\usepackage{makeidx}
\usepackage{epstopdf}

\begin{document}

\title{The Bass diffusion model on networks\\ 
with correlations and inhomogeneous advertising}

\author{M.L.\ Bertotti \footnote{Email address: marialetizia.bertotti@unibz.it}}
\affiliation{Free University of Bozen-Bolzano, Faculty of Science and Technology, Bolzano, Italy}
\author{J.\ Brunner \footnote{Email address: johannes.brunner@tis.bz.it}}
\affiliation{TIS Innovation Park, Bolzano, Italy}
\author{G.\ Modanese \footnote{Email address: giovanni.modanese@unibz.it}}
\affiliation{Free University of Bozen-Bolzano, Faculty of Science and Technology, Bolzano, Italy}

\linespread{0.9}

\begin{abstract}

The Bass model, which is an effective forecasting tool for innovation diffusion based on large collections of empirical data, assumes an homogeneous diffusion process. 
We introduce a network structure into this model and we investigate numerically the dynamics in the case of 
networks with link density $P(k)=c/k^\gamma$, where $k=1, \ldots , N$. 
The resulting curve of the total adoptions in time is qualitatively similar to the homogeneous Bass curve corresponding to a case with 
the same average number of connections. The peak of the adoptions, however, tends to occur earlier, particularly when $\gamma$ and $N$ are large (i.e., when there are few hubs with a large maximum number of  connections). Most interestingly, the adoption curve of the hubs anticipates the total adoption curve in a predictable way, with peak times which can be, for instance when $N=100$, between 10\% and  60\% of the total adoptions peak. This may allow to monitor the hubs for forecasting purposes. We also consider the case of networks with assortative and disassortative correlations and a case of inhomogeneous advertising where the publicity terms are ``targeted" on the hubs while maintaining their total cost constant. 

\end{abstract}

\maketitle

\section{Introduction}

The mathematical models which describe the diffusion of technological innovations belong to a category of diffusion and contagion models applicable to several phenomena and to different social structures. Frequent applications are, for instance, epidemics (of real diseases or computer viruses), spreading of information and rumors, opinion dynamics, marketing strategies for consumer products which take into account positive and negative word of mouth, etc.
 
These models typically consider a society composed by individuals who interact and communicate in various possible ways and can be in different ``states'' with respect to the diffusing phenomenon, while certain probabilities are assigned to pass from each state to another state. In the case of an epidemic, for instance, each individual can be assumed to be sound, infected or recovered/immunized; in the case of information diffusion each individual can be ignorant or informed, and informed people can act either as spreaders or as ``stiflers'' (persons who are informed but, for some reason, are not willing to pass the information on).

Usually the variables of the phenomenon are the normalized populations of the states, to be investigated as functions of time. The populations obey systems of differential equations; one is interested into the solutions of these equations at finite time (with characteristic evolution timescales which are often relevant to the problem), or into the ``asymptotic'' solutions at large time. Less frequently, periodic solutions are of interest, like for instance in the case of epidemics with cyclical outbursts. The differential equations are typically non-linear, because they need to include interaction terms describing the encounters between individuals, whose probabilities are proportional to both the populations of the respective states.

The main established tool for a mathematical description of the diffusion of technological innovations is the so-called Bass differential equation \cite{Bass}. Using this equation in conjunction with large empirical databases of statistical parameters, the experts of marketing and technology are able to make predictions about innovation processes in different kinds of markets. In particular, the diffusion rate of an innovation and the peak number of adoptions can be estimated with good accuracy.  The advantage of this method lies in its power and flexibility, also due to the fact that it is based on a differential equation and not on agent-based simulations. Also mathematical methods, however, need to be constantly improved, in order to reflect the evolving global features of the technology market, the development of new information channels and the fragmentation of society into several sectors which are more or less open to innovation. In fact, several improved versions of the Bass equation have been proposed \cite{Meade}.
 
The application of these techniques to the study of the diffusion of innovations has been historically accompanied by a wide qualitative and empirical literature, which includes subjects ranging from marketing strategy and industrial organization, to the sociological/medical studies concerning the diffusion of medical practices, contraception and prevention, introduction of innovative legislation, new didactic methods etc. \cite{Schi}. 
Classical works like those by Rogers \cite{Rogers} also contain ample reference to agricultural practices, diffusion of new sort of crops or farming machines, with adoption at the level of the single individuals or of entire organizations. Concepts which have been introduced more recently include the distinction between ``sustaining innovations'' and ``disruptive innovations'' \cite{Christensen}, and the idea of ``indigenous innovation'' \cite{Phelps} as compared to the science-driven innovation theorized by Schumpeter.

Already in the 1990's seminal papers like \cite{Abr} have highlighted the role of networks in the diffusion of innovation and information. These works, however, did not take advantage of the progress in network theory which took place a few years later. An important reference point for our work is the theoretical framework developed by Vespignani et al. for the statistical description of networks \cite{Bog1,Bog2}. These methods have been originally employed to analyze the diffusion of epidemics, and thus with a focus on different aspects, but then have found applications also for diffusion of information and innovation \cite{Boc}. 

A recent influential work is that by Watts and Dodds \cite{Wat}. They studied the effect on diffusion of a category of individuals called ``influencers'', who do not belong to the media business or other organizations, but are simply well-connected and able to influence (in some more or less well-defined sense) the choices of others. Watts and Dodds apply agent-based simulations running specific diffusion models (including the Bass model) to the structure of some real social networks. Our approach is different and more theoretical. We write (large) systems of differential equations which generalize the Bass model to various networks with power-law degree distribution, on the assumption, based on empirical work (see for instance Di Matteo et al. \cite{Gal} and the references therein), that the social networks involved in the diffusion of innovation are of this kind. 

In the next section we shall discuss the characterization of the diffusion curve as an ``S-curve'' with variable steepness. A general parametrization of S-curves has been given by Ghanbarnejad et al. in the context of linguistic innovation \cite{Ger}. Among the recent works on diffusion on networks, particularly significant are the exact solutions by Gleeson \cite{Gle} and the methods of McCullen et al. \cite{Mac}. Rahmandad and Sterman \cite{Ste} offer a detailed comparison between agent-based and differential equation models and argue that differential equation models are typically unsuitable for the representation of heterogeneous agents; our method represents a progress under this respect.

The outline of the paper is the following. In Sect.\ \ref{structure} we introduce our general framework for the formulation of the Bass equation on a generic network through the statistical link density $P(k)$ and the correlation functions $P(h|k)$. We define the partial cumulative adoption function referred to the single link classes and replace the homogeneous imitation term of the original Bass equation with a sum of terms describing the imitation between individuals belonging to different link classes. We take into account the correlation functions and the condition of constant average connectivity, in order to allow comparison with different networks 
and with the original Bass model (to which, for brevity, we also refer throughout the paper as to the homogeneous model). 
In Sect.\ \ref{steep} we introduce a parameter which measures the steepness of the S-curve $F(t)$ through the ratio between height and width at half-height of the graph of its derivative $f(t)$. In Sect.\ \ref{results} we discuss the results of numerical solutions with uncorrelated networks
(obtained with standard Runge-Kutta methods), in particular the dependence of the peak time $T$ on the power-law exponent $\gamma$ and its relation to the peak time $T_N$ of the most connected individuals. In Sect.\ \ref{AD} we consider networks with assortative and disassortative correlations. We develop a method for the explicit construction of the corresponding correlation matrices and discuss its applicability. Results of the numerical solutions are compared to those for uncorrelated networks. In Sect.\ \ref{pvariabile} we introduce link-dependent publicity terms $p_i$, weighted in such a way that the advertising is targeted on the hubs, with the same total cost. The results of the numerical simulations are compared to those with homogeneous advertising. Finally, Sect.\ \ref{Conc} comprises our conclusions and a summary of the main results.

\section{Network structure}
\label{structure}

The Bass equation has the form
\begin{equation}
\frac{{d{F}(t)}}{{dt}} = [1 - F(t)]\left[ {p + qF(t)} \right]
\end{equation}
where $F(t)$ is the cumulative fraction of adopters at the time $t$, measured in years. For instance, if at $t=4$ we have $F=0.7$, this means that after 4 years 70\% of the potential adopters have actually adopted the innovation. The function $F$ has a characteristic ``leaning S'' shape and approaches 1 at large times.

The derivative $f(t)=\frac{{d{F}(t)}}{{dt}}$ is the adoption probability in an interval of one year. Its graph exhibits a maximum called the peak of innovation diffusion. Producers of the innovation are clearly interested into forecasts of this peak.

The ``publicity'' parameter $p$ gives the probability per year of adoptions entirely due to the effect of advertising, while the ``imitation parameter'' $q$, multiplied by the fraction $F$ of actual adopters, gives the conditioned probability per year of adoptions due to word-of-mouth effects. In the example above, if $q=0.4$, in the 5th year the equation predicts the adoption rate due to imitation to be 8.4\% of the potential adopters.

A possible improvement of the Bass model is the introduction of classes of individuals who have more or less numerous links with others, with a corresponding dynamics on a network.

In order to introduce a network structure we proceed at first by analogy with the equations for the diffusion of epidemics and rumors. The first step is a statistical/probabilistic characterization of the network. This clearly has some limitations in comparison with a complete characterization through the adiacency matrix $a_{\mu \nu}$ (for which the indices run over all nodes of the network). We shall discuss later whether the information about the higher-order correlations contained in $a_{\mu \nu}$ can be inserted into the differential equations; until now, this has only been done for linear equations or for agent-based models. On the other hand, the statistical description of the network allows to summarize its mean features in a relatively small number of parameters, namely the link density $P(k)$ (the fraction of nodes having $k$ links, where $k=1 \ldots N$ and $N$ is the maximum number of links one node can have), 
and the correlation functions $P(k | i)$, which give the conditioned probability that an individual with $i$ links is connected to one with $k$ links. 

In the quadratic imitation term of the Bass equation, an homogeneous mixing is understood, i.e., each individual is supposed to have the same probability to meet any other individual. In the context of statistical mechanics, this is also called the ``molecular chaos hypothesis''. In a real society the probability of the encounters depends on the network of social relationships, which in turn defines the matrices $P(k|i)$. 
Therefore the imitation term will become in the $i$-th equation a sum of the form $ iq\sum\limits_{h = 1}^N {P(h|i){F_h}(t)}$, where we have supposed that the total population is subdivided into $N$ ``link classes'' indexed by $i=1,...,N$, with populations $F_i(t)$. In other words, $F_i(t)$ denotes the fraction of the total population composed by individuals with $i$ links to other individuals, who at the time $t$ have adopted the innovation. Admitting that, in the end, all individuals will adopt, we will have $F_i(0)=0$, $F_i(+\infty)=P(i)$. It is convenient to define the quantities $G_i(t)=F_i(t)/P(i)$, representing the fraction of potential adopters with $i$ links that at the time $t$ have actually adopted the innovation.
It is worth recalling here that the strategy of generalizing the evolution equations of some system by considering variables which group nodes with the same degree is referred to in the recent literature as ``heterogeneous mean-field approach" \cite{VespNaturePh}.

Let us give a simple numerical example. Suppose there are only 4 link classes, with number of links equal to 1, 2, 3, 4, respectively, and assume a link density of the form $P(i)=c/i$, where $c$ is a normalization constant ($c^{-1}=\sum\limits_{i = 1}^N i^{-1}$). It is immediate to see that $P(1)=0.48$, $P(2)=0.24$, $P(3)=0.16$ and $P(4)=0.12$.
Now, if at the time $t$ we have for instance $G_1(t)=0.5$, $G_2(t)=0.75$, this means that at that time 50\% of the individuals with 1 link (corresponding to 24\% of the total population) have adopted, and 75\% of the individuals with 2 links (corresponding to 18\% of the total population) have adopted. 

With these notations the network Bass equation becomes a system of non-linear first order differential equations
\begin{equation}
\frac{{d{G_i}(t)}}{{dt}} = [1 - {G_i}(t)]\left[ {p + iq\sum\limits_{h = 1}^N {P(h|i){G_h}(t)} } \right] \qquad i=1,...,N
\label{BassNet}
\end{equation}
with initial conditions $G_i(0)=0$. The factor $(1-G_i)$ on the r.h.s.\ shows that the asymptotic value of the functions $G_i(t)$ will be $G_i(+\infty)=1$, as expected. We can easily solve these equations numerically and plot the solutions, after fixing the link density $P(i)$ and the correlations $P(h|i)$. In the plots we re-introduce the quantities $F_i(t)$, $f_i(t)=F'_i(t)$, for a more direct comparison with the original Bass equation. 

Note that the matrices $P(i)$, $P(h|i)$ must obey the Network Closure Condition (NCC)
\begin{equation}
i P(k|i) P(i) = k P(i|k) P(k), \qquad \forall i ,k  = 1, \ldots ,N \ .
\label{cnc}
\end{equation}
Further conditions are normalizations $\sum\limits_{i = 1}^N P(i|k) =1$,  $\sum\limits_{k = 1}^N P(k) =1$ and positivity $P(i|k) \geq 0$, $P(i) \geq 0$. 

A first interesting check is the dependence of the diffusion phenomenon on the exponent  $\gamma$ for a network with link density $P(i)=c/i^\gamma$. This can be a meaningful check only if the average number of links per individual is kept fixed as  $\gamma$ varies. A proper question is, for instance: if in a certain network individuals have, on average, 2 links, is the diffusion faster when everyone has exactly or almost exactly 2 links, like in a random network with distribution sharply centered around $i$=2? Or is the diffusion faster when there are a few individuals with many more than 2 links, and a lot of individuals with just 1 link? 

Networks with link density of the form $P(i)=c/i^\gamma$, where $k=1, \ldots , N$
have been the object during the last few decades of intense investigation (see e.g. \cite{Bog2,Boc,Newman,NewmanSIAM}
and references therein).
Due to an invariance property of their
degree distribution, they are also known as scale-free networks, even if
this terminology is most frequently reserved for those 
of them
which have exponent in the range $2 < \gamma < 3$, which are frequently occurring in the real world.

The exponent $\gamma$ of a network with power-law degree distribution gives a measure of this inequality: when $\gamma$ is large, the number of hubs is very small. The average number of links per node in a network is
\begin{equation}
\langle k \rangle = \sum\limits_{k = 1}^N kP(k).
\end{equation}
Therefore, in order to compare the results obtained with different $\gamma$ we must normalize the $q$-coefficient in the network Bass equation (\ref{BassNet}) with this factor. The simplest way to assign the correlation matrix $P(h|i)$ is to consider at first uncorrelated networks, for which
\begin{equation} 
P(h|i) = \frac{h P(h)}{ \sum\limits_{k = 1}^N k P(k)} = \frac{c h^{1-\gamma}}{ \sum\limits_{k = 1}^N k P(k)} ,
\end{equation}
where $c^{-1}= \sum\limits_{k = 1}^N k^{-\gamma}$. Note that $P(h|i)$ is independent from $i$. This matrix $P(h|i)$ satisfies automatically the NCC condition for any $\gamma$. We shall consider later the case of correlated networks of the assortative or disassortative kind.

Now we solve the equation system (\ref{BassNet}) for different values of $\gamma$. We choose as single parameter characterizing the diffusion process the time $T$ of the peak in the diffusion rate (the peak of the sales, in marketing applications). This is the maximum of the function $f_{tot}= \sum\limits_{i = 1}^N f_i$ which gives the total rate of new adoptions summed over all link classes. 

\subsection{Steepness of the adoption peak}
\label{steep}

The Bass density function $f(t)$ describing the number of new adoptions in the unity of time is
\begin{equation} 
f(t) = \frac{{{\alpha ^2}}}{p}\frac{{{e^{ - \alpha t}}}}{{{{\left( {1 + \beta {e^{ - \alpha t}}} \right)}^2}}}
\end{equation}
where $\alpha=p+q$ and $\beta=q/p$. The adoption peak occurs at the time $T=\ln \beta / \alpha$, with $f(T)=\alpha^2/(4q)$. Denoting by $t_1$ and $t_2$ the instants when $f(t_1)=f(t_2)=\frac{1}{2} f(T)$, one obtains
\begin{equation} 
{t_2} - {t_1} = \frac{1}{\alpha }\ln \frac{{3 + 2\sqrt 2 }}{{3 - 2\sqrt 2 }} = \frac{1}{\alpha }\zeta .
\end{equation}
Let us denote by $\Delta$ the ratio between the height of the peak and its width at half-height. We find
\begin{equation} 
\Delta  = \frac{{f(T)}}{{{t_2} - {t_1}}} = \frac{{\alpha ^3}}{{{4\zeta q}}} \simeq \frac{{{(p + q)}^3}}{{{14.1q}}} .
\end{equation}
This ratio depends, of course, on the time unity (the standard unit is years, in the Bass model).

For comparison recall that, for instance, in the sequence of approximations of the Dirac delta
\begin{equation} 
{f_n}(x) = \frac{1}{\pi }\frac{n}{{1 + {{(nx)}^2}}}
\end{equation}
(derivatives of the smoothed step functions ${F_n}(x) = \frac{1}{\pi }\arctan (nx)$) one has $f_n(0)=n/\pi$, $x_{1,2}=\pm 1/n$ and therefore $\Delta=n^2/2\pi$.

In our case, the values of $T$, $f(T)$, $t_{1,2}$ are obtained numerically with the desired approximation by sampling the numerical solution for $f$.

\begin{figure}
\begin{center}
\includegraphics[width=10cm,height=6cm]{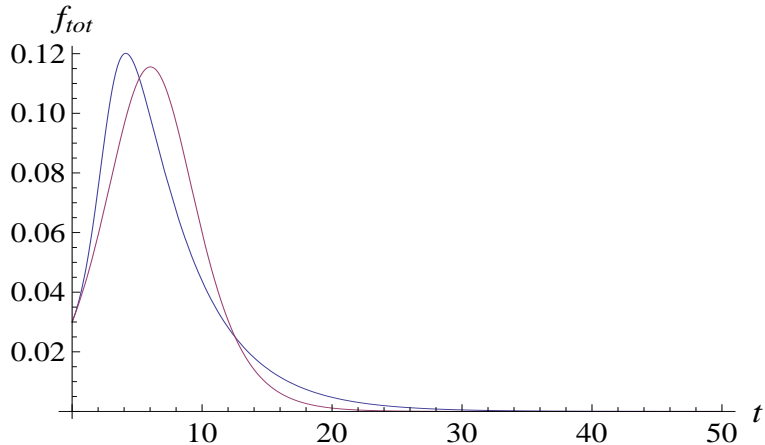}
\caption{Total adoptions $f_{tot}=\sum_{i=1}^N f_i$ as a function of time (blue curve), for an uncorrelated network with density proportional to $1/i^2$ ($\gamma=2$), with $N=15$. The violet curve is the simple Bass function with the same $q$ and $p$ ($q=0.4$, $p=0.03$). The function $f_{tot}$ peaks approx.\ at $t=4.1$, while the simple Bass function peaks approx.\ at $t=6.1$.} 
\label{f1}
\end{center}  
\end{figure}

\begin{figure}
\begin{center}
\includegraphics[width=10cm,height=6cm]{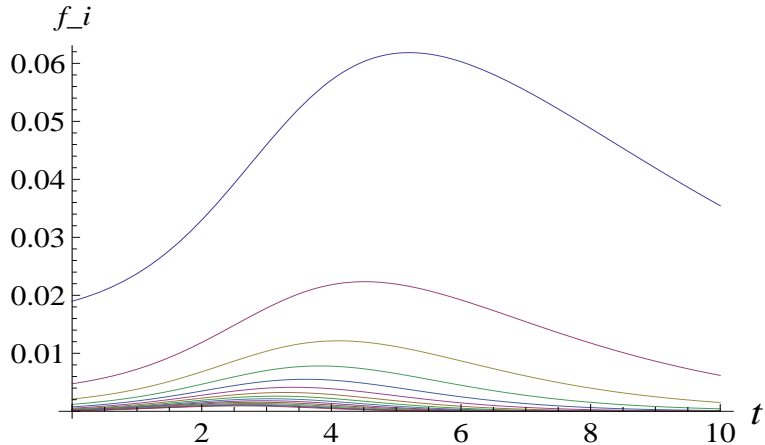}
\caption{Partial adoptions in time for the same parameters as in Fig.\ \ref{f1}. The function which has the largest value at $t=0$ is $f_1$, representing the adoptions of individuals with only 1 link. The function $f_{15}$, which represents the adoptions of the most connected individuals, peaks approx.\ at $t=2.5$.} 
\label{f2}
\end{center}  
\end{figure}

\subsection{Results for uncorrelated networks}
\label{results}

As shown in Tab.\ \ref{peaks}, the peak time $T$ of the new adoptions decreases when the exponent $\gamma$ grows, that is, when the network contains less hubs with more links. For $\gamma=2$, $T$ is about 2/3 of the diffusion peak time of the homogeneous Bass curve with the same average connectivity. Therefore the network structure has a considerable acceleration effect on the diffusion process. This effect appears to reverse, however, when $\gamma$ reaches the values 5/2. 

\begin{table}[h]
\begin{center}
\begin{tabular}{|c|c|c|c|}
\hline
$\gamma$  & \ $T$  \ & \ $T_{N}$  \ & \ $\Delta$  \nonumber \\
\hline
\ 1/4 \ & \ 5.1  \ & \ 4.4 \ & \ 0.0166  \nonumber \\
\hline
\ 1/2 \ & \ 4.9 \ & \ 4.1 \ & \ 0.0174  \nonumber \\
\hline
\ 3/4 \ & \ 4.6 \ & \ 3.8 \ & \ 0.0178  \nonumber \\
\hline
\ 1 \ & \ 4.4 \ & \  3.5 \ & \ 0.0184  \nonumber \\
\hline
\ 3/2 \ & \ 4.1 \ & \ 2.9 \ & \ 0.0182  \nonumber \\
\hline
\ 2 \ & \ 4.1 \ & \  2.5 \ & \  0.0172   \nonumber \\
\hline
\ 5/2 \ & \ 4.5 \ & \ 2.4 \ & \ 0.0159  \nonumber \\
\hline
\ 3 \ & \ 5.0 \ & \ 2.3 \ & \ 0.0151  \nonumber \\
\hline
\end{tabular}
\end{center}
\caption
{Time of adoption peak for different values of the power-law exponent $\gamma$ in the case of uncorrelated networks. $T$ is the total adoption time, $T_N$ is the partial adoption time of the most connected individuals, belonging to the class with $N$ links. Here $N=15$. $\Delta$ is the peak steepness, namely the ratio between its height and its width at half-height. The homogeneous Bass function with the same $q$ and $p$ ($q=0.4$, $p=0.03$) peaks approx.\ at $t=6.1$ and has steepness $\Delta=0.0141$.}
\label{peaks}
\end{table}

The shape of the total adoption rate $f_{tot}(t)$ (Fig.\ \ref{f1}) is not very different from the homogeneous Bass curve, though the peak is in general steeper. The steepness $\Delta$ attains a maximum for $\gamma=1$.

From the graphs of the partial adoption rates $f_{i}(t)$ (Fig.\ \ref{f2}), showing the adoption in the different link classes, we see that the most connected individuals adopt earlier; in the case $\gamma=2$, for instance, the adoption peak for individuals with 15 links occurs at $T_{15}=2.5$, while the peak of the total adoptions occurs at $T=4.1$ (Tab.\ \ref{peaks}). Therefore, monitoring the most connected individuals at the beginning of the diffusion process is a possible way to predict the occurrence of the peak of the total adoptions, even if one does not know in advance the $q$ parameter for the specific diffusion process considered. One needs, of course, some (statistical) information on the social network, in order to define the exponent $\gamma$ and the parameter $N$.

Another possibility is to apply the model to situations where the $q$ parameter is known from empirical data on the specific diffusion process, in order to predict the partial adoption peaks of different link categories.

\section{Assortative and disassortative case}
\label{AD}

Until now we have considered uncorrelated networks. It is known, 
however, that in many real situations the networks tend to be 
either assortative (A) or disassortative (D). This concept has been 
introduced by Newman \cite{Newman,NewmanSIAM}, who has explicitly constructed 
mathematical networks displaying these features and has analized 
some typical natural and social systems in which the correlations 
tend to be respectively of the A and D type. The role of 
correlations in the models of epidemics diffusion has been 
investigated by Boguna et al. \cite{Bog1}. The main conclusion is that 
the correlations do not alter the main feature of the diffusion 
on scale-free networks, i.e., the absence of an epidemic 
threshold. While in the study of epidemics diffusion the 
focus is on the threshold, in our case the focus is on the diffusion 
peak time $T$, on the maximum slope of the S-curve (or $\Delta$ 
parameter) and on the difference between the total time $T$ and the 
``partial" times $T_i$ of the link classes. In order to check the 
effects of A and D correlations on these quantities we need a 
more explicit representation of the A/D property than that used 
by Boguna et al. \cite{Bog1}. They first define the function $k_{nn}(k)= 
\sum_h hP(h|k)$ which has the property of being increasing in the 
A case and decreasing in the D case; then they cast the diffusion 
equations into a form where only $k_{nn}(k)$ appears, and not the 
full correlation matrix $P(h|k)$. This is not possible in our 
equations, so in order to obtain numerical solutions we need to 
insert explicitly the correlation matrices. 

\subsection{Construction of the correlation matrices}

The construction of 
such matrices for any dimension is not trivial, because they must 
satisfy the positivity and Network Closure conditions (\ref{cnc}). We have 
developed a method which is helpful in most of the cases of 
interest. 

\smallskip

Let us first illustrate it in a simple case, namely 
that of power-law exponent $\gamma=1$, corresponding to a link 
density of the form $P(k)=c/k$. In this case it follows 
immediately from (\ref{cnc}) that $P(h|k)$ is symmetric. 

Then, in the D case we start and define
\begin{equation}
P_D(h|i)= |h-i|^\alpha, \qquad \alpha >0, \qquad h,i=1,\ldots ,N.
\label{P_D}
\end{equation}
The idea is to obtain matrix elements which are zero on the 
diagonal and grow when one moves away from the diagonal; this is 
actually the defining feature of D correlations.

Next we must normalize each column of the matrix to 1 in order to respect the property $\sum_h P(h|i)=1$ for any $i$. 
To this end, we compute the sums $C_i = \sum_h P(h|i)$. For the matrix $P_D$ defined in (\ref{P_D}), 
it results
$$
C_1 = C_N > C_j \qquad     \forall j = 2, \ldots, N-1 .
$$
Indeed, $C_N = C_1$ in view of the symmetry of the matrix. To see that $C_1 > C_j$ for all $j = 2, \ldots, N-1$, we split each of the expressions of $C_1$ and $C_j$
in two sums, one containing $j - 1$ terms and the other containing $N - j$ terms. We get
\begin{equation}
C_1 =  \sum_{h=2}^{N} \, |h - 1|^{\alpha} = \sum_{i=1}^{j-1} \, |i|^{\alpha} + \sum_{i=j}^{N-1} \, |i|^{\alpha} ,
\label{C1sumoftwopieces}
\end{equation}
whereas
\begin{equation}
C_j = \sum_{h=1}^{j-1} \, |h - j|^{\alpha} + \sum_{h=j+1}^{N} \, |h-j|^{\alpha}  \qquad     \hbox{for} \ j = 2, \ldots, N-1 .
\label{Cjsumoftwopieces}
\end{equation}
The first of the two sums on the right hand side in $(\ref{Cjsumoftwopieces})$ is in fact equal to
$$
\sum_{h=1}^{j-1} \, |j - h|^{\alpha} = \sum_{j-h=j-1}^{1} \, |j - h|^{\alpha} = \sum_{i=1}^{j-1} \, |i|^{\alpha} ,
$$
and is hence equal to the first sum on the right hand side in $(\ref{C1sumoftwopieces})$.
Also, since here $j > 1$ and consequently $ i + 1 - j < i$, and $0 <  i +1 -j$ for $i \ge j$, the second sum on the right hand side in $(\ref{Cjsumoftwopieces})$ is equal to
$$
\sum_{h-1=j}^{N-1} \, |h-j|^{\alpha} = \sum_{i=j}^{N-1} \, |i +1 -j|^{\alpha} = \sum_{i=j}^{N-1} \, (i +1 -j)^{\alpha} < \sum_{i=j}^{N-1} \, (i)^{\alpha}  = \sum_{i=j}^{N-1} \, |i|^{\alpha} ,
$$
and is hence strictly less than the second sum on the right hand side in $(\ref{C1sumoftwopieces})$.

Then we can re-define the diagonal elements in such a way that the new column sums $C_i'$ are equal to $C_1$ and $C_N$ for any $i$, namely we set
\begin{equation}
P_D'(i|i)= C_1-C_i, \qquad i=1,\ldots ,N
\label{P_Dprime}
\end{equation}
and keep $P_D'(i|j) = P_D(i|j)$ for $i \ne j$,
so that
\begin{equation}
C_i'=C_i-|i-i|^\alpha+C_1-C_i=C_1, \qquad i=1,\ldots ,N .
\label{Ci-prime}
\end{equation}
This allows a last step in which we normalize the entire matrix with the same factor $C_1$, maintaining its symmetry property:
\begin{equation}
P_D''(h|i)= C_1^{-1} P_D'(h|i), \qquad i,h=1,\ldots ,N .
\label{P_Dsecond}
\end{equation}
An undesired consequence of the re-definition of the terms $P_D(i|i)$ is that some new diagonal elements actually become larger 
than the closest non-diagonal elements. Nevertheless,  at least when $\alpha = 1$, it can be shown that
the function $k_{nn}(k) = \sum_h hP_D''(h|k)$ is decreasing and hence
the matrix is disassortative. We postpone this calculation in an Appendix.

In the A case we want to obtain a matrix in which the elements on the diagonal are the largest, and then decrease when we move away from the diagonal. A suitable symmetric expression has the form
\begin{equation}
{P_A}(h|i) = \left\{ \begin{array}{l}
|h - i{|^{ - \alpha }} \qquad {\rm if} \qquad h \ne i,   \qquad \alpha > 0 , \\
1 \qquad \qquad \ \ \ \  {\rm if} \qquad h = i.
\end{array} \right.
\end{equation}
We then proceed in a similar way as for the D case. Here, however, the column sums $C_i$ attain greater values in correspondence to central columns. 
It is therefore convenient to consider only matrices with dimension $N$ odd, such that $N_1=(N+1)/2$ is integer, and re-define the diagonal elements as
\begin{equation}
P_A'(i|i)= C_{N_1}-C_i, \qquad i=1,\ldots ,N ,
\label{P_Aprime}
\end{equation}
while keeping $P_A'(i|j) = P_A(i|j)$ for $i \ne j$.
Finally, the whole matrix is renormalized by one single factor as
\begin{equation}
P_A''(h|i)= C_{N_1}^{-1} P_A'(h|i), \qquad i,h=1,\ldots ,N .
\label{P_Asecond}
\end{equation}
Arguing in a similar way as for the D case, at least when $\alpha = 1$, one can prove that the function $k_{nn}(k)$
increases as $k$ increases. Technical details can be found in the Appendix

\smallskip

Let us now turn to the general case of a network with generic power-law exponent $\gamma$, i.e., with link density $P(k)$ proportional to $k^{-\gamma}$. In this case the NCC condition implies that the correlation matrix is not just symmetric, but such that
\begin{equation}
P(h|i) = P(i|h)\frac{{{h^{1 - \gamma }}}}{{{i^{1 - \gamma }}}}.
\label{Pnonsimm}
\end{equation}

For simplicity we consider $\gamma > 1$ and we define the elements on the diagonal or above it, i.e.\ with $h \leq i$, as follows. For the D case:
%\begin{equation}
%P_D(h|i)= |h-i|^\alpha, \qquad \alpha >0, \qquad h \leq i, \qquad h,i=1,\ldots ,N.
%\label{P_Dnonsimm}
%\end{equation}
\begin{equation}
P_D(h|i)= |h-i|,  \qquad h \leq i, \qquad h,i=1,\ldots ,N.
\label{P_Dnonsimm}
\end{equation}
For the A case:
\begin{equation}
{P_A}(h|i) = \left\{ {\begin{array}{*{20}{l}}
{|h - i{|^{ - 1}}\quad \;\;{\kern 1pt} {\rm{if}}\quad \;\;\;{\kern 1pt} h < i},     \\
{1\quad \;\;\;{\kern 1pt} \quad \;\;\;{\kern 1pt} \;\;\;\;{\rm{if}}\quad \;\;\;{\kern 1pt} h = i.}
\end{array}} \right.
\label{P_Anonsimm}
\end{equation}
The elements under the diagonal, i.e.\ with $h>i$, are computed according to (\ref{Pnonsimm}).

Then we define the column sums $C_i$ like above and proceed in a similar way. 
It turns out here that, leading to too large values on the diagonal, 
and due to difficulties related to the asymmetry,
the present procedure is not suitable in general
(at least without further adjustments)
for the construction of D matrices when $\gamma \neq 1$. The challenge of constructing explicitly 
the D matrices for $\gamma \neq 1$ thus remains open.
Nonetheless, the procedure works for A matrices as outlined below.
In practice, for the purposes of this work it is not restrictive to limit ourselves to A matrices. Indeed, it is well know that 
%in the social networks 
%involved in the diffusion of innovation or information 
``essentially all social networks measured appear to be assortative" \cite{NewmanSIAM},
thus also in social networks involved in the diffusion of innovation
%and then also the social networks involved in the diffusion of innovation.
the A-correlations are predominant.

Coming back to the construction of an assortative matrix when 
%$\gamma \ne 1$ and more properly when 
$\gamma > 1$,
a little thought can convince the reader that the maximum column sum isn't  attained now in correspondence to 
the first or the last or the central column, but 
in correspondence to an intermediate
column, say the $j^*$-th one with $1 \le j^* \le N$, i.e.
\begin{equation}
C_{j^*} = \max_{j=1,...N} \{C_j\} .
\label{Cmax}
\end{equation}
We then re-define the diagonal elements as
\begin{equation}
P_A'(i|i)= C_{j^*}-C_i, \qquad i=1,\ldots ,N ,
\label{P_Aprimeggreaterthan1}
\end{equation}
while keeping $P_A'(i|j) = P_A(i|j)$ for $i \ne j$, and we renormalize the whole matrix by one single factor as
\begin{equation}
P_A''(h|i)= C_{j^*}^{-1} P_A'(h|i), \qquad i,h=1,\ldots ,N .
\label{P_Asecondggreaterthan1}
\end{equation}
We prove in the Appendix that the function $k_{nn}(k)$
increases as $k$ increases. 

\smallskip

Table \ref{peaksA} summarizes the results obtained for the diffusion peak time $T$ and for the ratio $\Delta$ between the peak height and its half-height width. In the D case with $\gamma =1$, $T$ is larger than in the uncorrelated case; thus the adoption peak occurs later, but then adoption is faster, with a slope ratio $\Delta=0.0186$, which is the largest value of all those considered (excluding the case of targeted advertising -- compare Sect.\ \ref{pvariabile}). For the A case, let us start from $\gamma =1$ and increase $\gamma$ (thus in the direction of having fewer hubs, but each with more connections, at constant average connectivity). When $\gamma =3/2$ we observe a diminution of $T$ and also of $\Delta$, that is, a diffusion which is faster and less abrupt; the trend is confirmed for $\gamma =2$. 

\begin{table}[h]
\begin{center}
\begin{tabular}{|c|c|c|c|c|}
\hline
$\gamma$  & \ $T$ (Assort.) \ & \ $\Delta_A$  \ & \ $T$ (Disass.) & \ $\Delta_D$ \ \nonumber \\
\hline
\ 1 \ & \ 3.9 \ & \ 0.0161 \ & \ 5.2 \ & 0.0163 \nonumber \\
\hline
\ 3/2 \ & \ 3.3 \ & \  0.0132 \ & \  \ &  \nonumber \\
\hline
\ 2 \ & \ 3.2 \ & \  0.0100 \ & \  \ &   \nonumber \\
\hline
\end{tabular}
\end{center}
\caption
{Time of adoption peak for different values of the power-law exponent $\gamma$ in the presence of assortative and disassortative correlations. ($N=15$, $\alpha=1$.) The simple Bass function with the same $q$ and $p$ ($q=0.4$, $p=0.03$) peaks approx.\ at $t=6.1$ and has steepness $\Delta=0.0141$. For assortative networks in the case $N=100$, $2 \le \lambda \le 3$ see Fig.\ \ref{T_of-gamma}.}
\label{peaksA}
\end{table}

\section{Link-dependent publicity term}
\label{pvariabile}

Until now we have supposed that the publicity term $p$ is the same for the whole population, and we have taken the network structure into account only in the imitation term. It is however clear, in view of the results described so far, that the hubs of a network with power-law degree distribution have a crucial role in the diffusion process. Therefore, fixed a certain total amount available for publicity, it would probably be more effective to address the advertisement to the hubs (also called by Rogers ``the opinion leaders'') rather than to poorly connected individuals. Since the individuals with $k$ links are $k^\gamma$ times less numerous than those with one link, if the $p$ coefficient (supposed proportional to the expenditure for publicity) is $k^\gamma$  times larger for the former than for the latter, then the total cost of publicity will be the same. Let us denote by $p_k$ the publicity coefficient for the link class $k$ and re-write the equations (\ref{BassNet}) as
\begin{equation}
\frac{{d{G_i}(t)}}{{dt}} = [1 - {G_i}(t)]\left[ {p_i + iq\sum\limits_{h = 1}^N {P(h|i){G_h}(t)} } \right] \qquad i=1,...,N
\label{BassNet-pvar}
\end{equation}
where we set $p_k=c_1 k^\gamma p$, being $c_1$ an appropriate normalization constant determined through the condition $\sum_k p_k P(k)=p$, whence $c_1=1/(cN)$.

\begin{figure}
\begin{center}
\includegraphics[width=10cm,height=6cm]{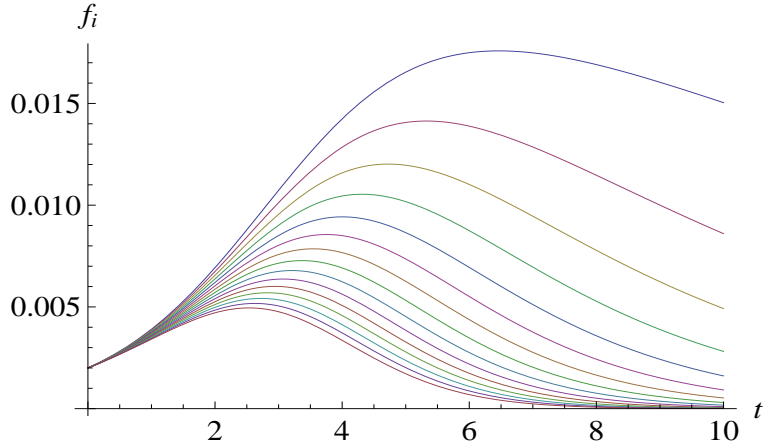}
\caption{Partial adoptions in time for an uncorrelated network with density proportional to $1/i$ ($\gamma=1$) and $N=15$. Here the publicity coefficients $p_i$ are weighted in inverse proportion to the population, thus giving the same total advertising cost but accelerated adoption from the hubs (lower curves); this is apparent from the fact that the initial diffusion rate is exactly the same for all link classes.} 
\label{p-var}
\end{center}  
\end{figure}

\begin{figure}
\begin{center}
\includegraphics[width=13cm,height=5cm]{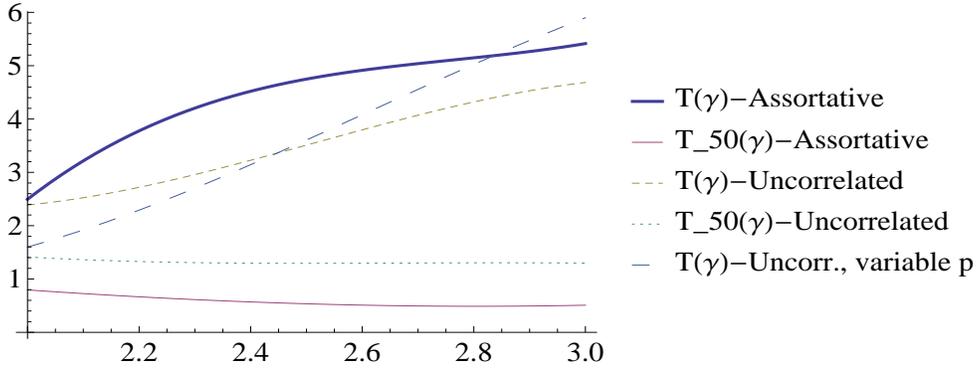}
\caption{Dependence on the scale-free exponent $\gamma$ of the total diffusion peak time $T$ and of the partial diffusion peak time $T_{50}$ (for individuals with 50 links) in a network with largest degree $N=100$, in the uncorrelated and assortative case. The peak time $T_{50}$ has been chosen as indicator of the anticipated diffusion in the hubs, instead of $T_{100}$, because in these networks the hubs with 100 links are a very small fraction of the total population, and thus are not significant for statistical monitoring. Nevertheless, the anticipation effect is clearly very strong, especially for the assortative networks: for $\gamma=2$, $T_{50}$ is approx. 20\% of $T$, and for $\gamma=3$ it is approx. 10 \% of $T$. Note that for uncorrelated networks the total diffusion time is smaller, but the anticipation effect is weaker. The dependence $T(\gamma)$ in the case of variable $p$ coefficients is also shown.} 
\label{T_of-gamma}
\end{center}  
\end{figure}

The numerical solutions of these equations show that diffusion indeed becomes still faster, both for the total population and for the hubs. See Fig.\ \ref{p-var}, where it is also apparent how the initial diffusion rate is exactly the same for all link classes, since the classes with more links are less populated but adopt quicker, in inverse proportion to their population. The adoption peak times $T$ and steepness parameters $\Delta$ in correspondence of different power-law exponents are given in Tab.\ \ref{peaks2}. These data are obtained with uncorrelated networks, so they should be compared to those of Tab.\  \ref{peaks}.
Results of the numerical solutions with large networks ($N=100$) in the scale-free range $2 \leq \gamma \leq 3$ are given in Fig.\ \ref{T_of-gamma} for comparison with the other cases. The plot of $T(\gamma)$ shows that diffusion is markedly faster in the variable-$p$ case when $\gamma$ is close to 2, but the advantage is lost when 
$\gamma$ approaches 3. Also, the anticipation effect is absent in the range $2 \leq \gamma \leq 3$.

\begin{table}[h]
\begin{center}
\begin{tabular}{|c|c|c|c|c|}
\hline
$\gamma$  & \ $T$  \ & \ $T_{N}$  \ & \ $\Delta$  \nonumber \\
\hline
\ 1/4 \ & \ 4.9  \ & \ 4.1 \ & \ 0.0168  \nonumber \\
\hline
\ 1/2 \ & \ 4.5 \ & \ 3.6 \ & \ 0.0172  \nonumber \\
\hline
\ 3/4 \ & \ 4.1 \ & \ 3.1 \ & \ 0.0176  \nonumber \\
\hline
\ 1 \ & \ 3.7 \ & \  2.5 \ & \ 0.0178  \nonumber \\
\hline
\ 3/2 \ & \ 3.1 \ & \ 1.2 \ & \ 0.0163  \nonumber \\
\hline
\ 2 \ & \ 3.1 \ & \  no peak \ & \  0.0137   \nonumber \\
\hline
\end{tabular}
\end{center}
\caption
{Time of adoption peak for different values of the power-law exponent $\gamma$ in the case of uncorrelated networks, with variable publicity coefficients $p_k$ weighted as $p_k=c_1 k^\gamma p$, thus giving the same total advertising cost but accelerated adoption from the hubs. $T$ is the total adoption time, $T_N$ is the partial adoption time of the most connected individuals, belonging to the class with $N$ links. Here $N=15$. $\Delta$ is the peak steepness, namely the ratio between its height and its width at half-height. The homogeneous Bass function with the same $q$ and $p$ ($q=0.4$, $p=0.03$) peaks approx.\ at $t=6.1$ and has steepness $\Delta=0.0141$. For results with $N = 100$ and $\gamma$ in the range $2 \le \gamma \le 3$ see Fig. \ \ref{T_of-gamma}.}
\label{peaks2}
\end{table}

Note that although the criterion of inverse proportionality $p_k=c_1 k^\gamma p$ allows to leave the total advertising cost unchanged, this is not the only possible choice. One could as well look for strategies allowing a diminution of advertising cost with an acceptable slowing down of diffusion, or viceversa, in some cases it may be preferable to spend more in advertising in order to obtain a faster diffusion. The advantage of an analytical model is that these parameters can easily be tuned and their effect immediately checked, 
see in this connection the Fig.\ \ref{T_of-gamma}. Also the parameters describing the diffusion process can be chosen in a different way; the parameters $T$ and $\Delta$ can be replaced by other measures of diffusion, like for instance diffusion percentiles etc.
 
\section{Conclusions}
\label{Conc}

In this work we have developed a network version of the Bass diffusion equation in the statistical network formalism introduced by Boguna et al. \cite{Bog1} for the study of epidemics on scale-free networks. The main technical aspects of our approach are the following:
\begin{enumerate}
\item
an appropriate definition of link classes and relative adoption rates, with normalization of the average connectivity for comparison to the homogeneous Bass equation;
\item
the definition of a measure of the steepness of the S-curve of diffusion, to be employed in conjunction with the diffusion peak time in the analysis of numerical solutions;
\item
a generalization of the imitation term, obtained through the consideration of the probabilities of encounters between members of the link classes, expressed by the correlation coefficients $P(h|k)$;
\item 
the explicit construction of correlation matrices of the assortative and disassortative type in any dimension, addressed to allow comparisons between the assortative, disassortative and uncorrelated case;
\item 
a generalization of the publicity term achieved by considering the possibility of a ``weighted'' advertising campaign aimed at accelerating diffusion among the hubs and consequently among the entire population.
\end{enumerate}

The main results which can be inferred from the numerical solutions of the model are the following:

\begin{enumerate}
\item
In the uncorrelated case, the peak time of the total adoptions decreases steadily when the power-law exponent $\gamma$ of the network grows from 1/4 (almost-homogeneous network) to 2 (network with few, much connected hubs); then the peak time grows in the interval $2 \le  \gamma \le 3$.
\item
The peak time in the adoptions of the hubs decreases in the whole interval $1/4 \le  \gamma \le 3$. This means that by monitoring the hubs it is always possible, in principle, to predict the peak of the total adoptions. Also, this can be done more effectively when the hubs are few and much connected. The prediction requires the numerical solution of the differential equations and the previous knowledge of the parameter $\gamma$; this is, however, a statistical parameter, which can be estimated without a complete mapping of the network.
\item
The presence of assortative correlations may accelerate diffusion, while the presence of disassortative correlations, where applicable,
slows it down. Specifically, with reference to large networks ($N=100$) in the scale-free range $2 \leq \gamma \leq 3$ (Fig.\ \ref{T_of-gamma}), the numerical solutions show that the total diffusion time is larger (with respect to uncorrelated networks), but the anticipation effect is stronger.
\item
The presence of advertising targeted on the hubs may further accelerate diffusion and improve the monitoring functions of the hubs. For instance,
with $\gamma=3/2$ the diffusion peak time of the hubs is less than 40\% of the total peak time (in a case with $N=15$, $q=0.4$, $p=0.03$).
However, for large networks in the scale-free range $2 \leq \gamma \leq 3$ the ``anticipation" effect of the hubs is absent and when $\gamma$ approaches 3, targeted advertising becomes less efficient than homogeneous advertising (Fig.\ \ref{T_of-gamma}).
\end{enumerate}

Finally, we point out that the best set of real data presently available for a comparison with our model is reported by Goldenberg et al. \cite{Gold}
who have analyzed the diffusion of some online services in a large Korean social network with $22$ million users. 
After identifying the network hubs and finding a scale-free degree distribution, the authors of \cite{Gold} measured several quantities 
related to the total speed of the adoption process, to hub and non-hub adoption timing and to hub adoption as predictor of total adoption timing.
The results are generally in agreement with the dynamics predicted by our model. 
A simple regression analysis and comparison with the homogeneous Bass model is also given in \cite{Gold}. 
The main focus of the work, however, is on marketing aspects. 
A more detailed comparison of our model with real data would require a dedicated data collection,
which is among our future plans.

\section{Appendix}
\label{App}

{\bf{I. Decreasing character of the function $k_{nn}(k) = \sum_h hP_D''(h|k)$ constructed in Section \ref{AD} 
in the $D$ case with $\gamma = 1$ (procedure with $\alpha = 1$)}}

Proof: 
We first rewrite for a generic $k$ the expression of $k_{nn}(k)$ as follows:
\begin{eqnarray*}
\sum_h hP_D''(h|k) = \frac{1}{C_1} \, \bigg[ k \,(C_1 - C_k) + \sum_{h=1}^{k-1} \, h \,(k - h)^{\alpha}  + \sum_{h=1}^{N-k} \, (h+k) \,h^{\alpha} \bigg] ,
\end{eqnarray*}
where
\begin{eqnarray*}
C_k &=& \sum_{h=1}^{k-1} \, (k - h)^{\alpha}  + \sum_{h=k+1}^{N} \, (h-k)^{\alpha} = \sum_{h=1}^{k-1} \, (k - h)^{\alpha}  + \sum_{h=k+1}^{N} \, (h-k)^{\alpha} \\ 
&=&  \sum_{h=1}^{k-1} \, (k - h)^{\alpha}  + \sum_{h-k=1}^{N-k} \, (h-k)^{\alpha}  = \sum_{h=1}^{k-1} \, (k - h)^{\alpha}  + \sum_{j=1}^{N-k} \, j^{\alpha} . 
\end{eqnarray*}
Accordingly, it is
\begin{eqnarray*}
k_{nn}(k) &=& 
\frac{1}{C_1} \, \bigg[ k \, C_1 - k \, \sum_{h=1}^{k-1} \, (k - h)^{\alpha} - k \, \sum_{j=1}^{N-k} \, j^{\alpha} \\
\qquad & & +\sum_{h=1}^{k-1} \, h \,(k - h)^{\alpha}  + \sum_{h=1}^{N-k} \, (h+k) \,h^{\alpha} \bigg] \\ 
&=& 
\frac{1}{C_1} \, \bigg[ k \, C_1 - \sum_{h=1}^{k-1} \, (k - h)^{\alpha + 1} - k \, \sum_{j=1}^{N-k} \, j^{\alpha} + \sum_{h=1}^{N-k} \, (h+k) \,h^{\alpha} \bigg] \\ 
&=& 
\frac{1}{C_1} \, \bigg[ k \, \sum_{j=1}^{N-1} \, j^{\alpha} - \sum_{h=1}^{k-1} \, (k - h)^{\alpha + 1} - k \, \sum_{j=1}^{N-k} \, j^{\alpha} + \sum_{h=1}^{N-k} \, (h+k) \,h^{\alpha} \bigg] \\ 
&=& 
\frac{1}{C_1} \, \bigg[ k \, \sum_{j=N-k+1}^{N-1} \, j^{\alpha} - \sum_{h=1}^{k-1} \, (k - h)^{\alpha + 1}  + \sum_{h=1}^{N-k} \, (h+k) \,h^{\alpha} \bigg] .
\end{eqnarray*}
In particular, if $\alpha = 1$, 
\begin{eqnarray*}
k_{nn}(k) = 
\frac{1}{C_1} \, \bigg[ k \, \sum_{j=N-k+1}^{N-1} \, j - \sum_{h=1}^{k-1} \, (k - h)^{2}  + \sum_{h=1}^{N-k} \, (h+k) \,h \bigg] .
\end{eqnarray*}
Since
\begin{eqnarray*}
\sum_{j=1}^n \, j = \frac{n\, (n+1)}{2} \quad \hbox{and} \quad \sum_{j=1}^n \, j^2 = \frac{n\, (n+1) \, (2n + 1)}{6} ,
\end{eqnarray*}
we then have
\begin{eqnarray*}
k_{nn}(k) &=& \frac{1}{C_1} \, \bigg[ \bigg( \frac{N (N-1)}{2} - \frac{(N-k) (N-k+1)}{2} \bigg) \, k -  \frac{(k-1) k (2k-1)}{6} \\
& &
\qquad +  \frac{(N-k)(N-k+1)(2N-2k+1)}{6} +  \frac{k (N-k)(N-k+1)}{2} \bigg] \\ 
&=& \frac{1}{C_1} \, \bigg[ \frac{N (N-1)}{2} \, k -  \frac{(k-1) k (2k-1)}{6} \\
& &
\qquad +  \frac{(N-k)(N-k+1)(2N-2k+1)}{6} \bigg] \\
&=& - \frac{2}{3} \, k^3 + (1 + N) \, k^2 - \frac{(2 + 9 N + 3 N^2)}{6} \, k +   \frac{(1 + 3 N + 2 N^2) \, N}{6}  . 
\end{eqnarray*}
The fact that $k_{nn}(k)$ decreases as $k$ increases can now be seen by evaluating the
derivative with respect to $x$ of the function $k_{nn}(x)$. It can be easily checked that it is always negative,
%(it is equal to $- 2 x^2 + (1 + N) \, x - \frac{(2 + 9 N + 3 N^2)}{6}$).
which proves the claim.

\medskip

{\bf{II. Increasing character of the function $k_{nn}(k) = \sum_h hP_A''(h|k)$ constructed in Section \ref{AD} in the $A$ case with $\gamma = 1$ 
(procedure with $\alpha = 1$)}}

Proof: 
We have for a generic $k$,
\begin{eqnarray*}
k_{nn}(k) &=& \sum_h hP_A''(h|k) \\
& = &
\frac{1}{C_{N_1}} \, \bigg[ k \,(C_{N_1} - C_k) + \sum_{h=1}^{k-1} \, h \,(k - h)^{-\alpha}  + \sum_{h=k+1}^{N} \, h \,(h - k)^{-\alpha}  \bigg] \\
& = &
\frac{1}{C_{N_1}} \, \bigg[ k \,(C_{N_1} - C_k) + \sum_{h=1}^{k-1} \, h \,(k - h)^{-\alpha}  + \sum_{j=1}^{N-k} \, (k+j) \,j^{-\alpha}  \bigg] ,
\end{eqnarray*}
where
\begin{eqnarray*}
C_k &=& \sum_{h=1}^{k-1} \, (k - h)^{-\alpha}  + 1 + \sum_{h=k+1}^{N} \, (h-k)^{-\alpha} = \sum_{h=1}^{k-1} \, (k - h)^{-\alpha} + 1 + \sum_{j=1}^{N-k} \, j^{-\alpha} .
%&=&  \sum_{h=1}^{k-1} \, (k - h)^{\alpha}  + \sum_{h-k=1}^{N-k} \, (h-k)^{\alpha}  = \sum_{h=1}^{k-1} \, (k - h)^{\alpha}  + \sum_{j=1}^{N-k} \, j^{\alpha} . 
\end{eqnarray*}
Hence,
\begin{eqnarray*}
k_{nn}(k) &=& 
\frac{1}{C_{N_1}} \, \bigg[ k \,C_{N_1} 
- k \, \sum_{h=1}^{k-1} \, (k - h)^{-\alpha}
- k 
- k \, \sum_{j=1}^{N-k} \, j^{-\alpha} \\
\qquad & & + \sum_{h=1}^{k-1} \, h \,(k - h)^{-\alpha}  + \sum_{j=1}^{N-k} \, (k+j) \,j^{-\alpha}  \bigg] \\
&=& 
\frac{1}{C_{N_1}} \, 
\bigg[ k \,(C_{N_1} - 1) - \sum_{h=1}^{k-1} \, (k - h)^{-(\alpha - 1)} - \sum_{j=1}^{N-k} \, j^{-(\alpha - 1)} \bigg] .
\end{eqnarray*}
In particular, if $\alpha = 1$, 
\begin{eqnarray*}
k_{nn}(k) &=& 
\frac{1}{C_{N_1}} \, 
\bigg[ k \,(C_{N_1} - 1) - (k - 1) + (N-k) \bigg] 
%\\
%&=& 
=
\frac{1}{C_{N_1}} \, 
\bigg[ k \,(C_{N_1} - 3) + 1 + N \bigg]  .
\end{eqnarray*}
Now, $C_{N_1} - 3 > 0$ provided $N \ge 5$ and this is certainly true for the networks we are considering.
It is then proved that $k_{nn}(k)$ increases as $k$ increases
and this proves the claim.

\medskip

{\bf{III. Increasing character of the function $k_{nn}(k) = \sum_h hP_A''(h|k)$ constructed in Section \ref{AD} in the $A$ case
with $\gamma > 1$}}

Proof: 
Recalling the definition of $C_{j^*} $ in (\ref{Cmax}) and taking since the beginning $\alpha = 1$, we have for a generic $k$,
\begin{eqnarray*}
k_{nn}(k) &=& \sum_h hP_A''(h|k) \\
& = &
\frac{1}{C_{j^*}} \, \bigg[ k \,(C_{j^*} - C_k) 
+ \sum_{h=1}^{k-1} \, h \,\frac{1}{(k - h)}  
+ \sum_{h=k+1}^{N} \, h \,\frac{1}{(h - k)} \, \bigg(\frac{k}{h}\bigg)^{\gamma - 1}  \bigg] ,
\end{eqnarray*}
where
\begin{eqnarray*}
C_k &=& \sum_{h=1}^{k-1} \, \frac{1}{(k - h)} + 1 + \sum_{h=k+1}^{N} \, \frac{1}{(h - k)} \, \bigg(\frac{k}{h}\bigg)^{\gamma - 1}  .
\end{eqnarray*}
Hence,
\begin{eqnarray*}
k_{nn}(k) &=& 
\frac{1}{C_{j^*}} \, \bigg[ (C_{j^*} - 1) \, k - (k - 1) +  \sum_{h=k+1}^{N} \, \bigg(\frac{k}{h}\bigg)^{\gamma - 1} \bigg] \\
&=&
\frac{1}{C_{j^*}} \, \bigg[ (C_{j^*} - 2) \, k + 1 +  \sum_{h=k+1}^{N} \, \bigg(\frac{k}{h}\bigg)^{\gamma - 1} \bigg]  .
\end{eqnarray*}
Differently from what done above, we cannot take any derivative here and hence we try and see whether
\begin{eqnarray}
k_{nn}(k) < k_{nn}(k+1) ,
\label{k<k+1}
\end{eqnarray}
which would guarantee the claimed increasing character of the function $k_{nn}$.
Equivalently, we try and see whether
\begin{eqnarray}
\sum_{h=k+1}^{N} \, \bigg(\frac{k}{h}\bigg)^{\gamma - 1}
<
(C_{j^*} - 2) +  \sum_{h=k+2}^{N} \, \bigg(\frac{k+1}{h}\bigg)^{\gamma - 1}  .
\label{wanted}
\end{eqnarray}
The left hand side in (\ref{wanted}) is the sum of the $N-k$ terms
\begin{eqnarray*}
\bigg(\frac{k}{k+1}\bigg)^{\gamma - 1} + \bigg(\frac{k}{k+2}\bigg)^{\gamma - 1} + \ ... \ + \bigg(\frac{k}{N}\bigg)^{\gamma - 1} ,
%\label{first sum}
\end{eqnarray*}
whereas
the sum in the right hand side in (\ref{wanted}) is the sum of the $N-k-1$ terms
\begin{eqnarray*}
\bigg(\frac{k+1}{k+2}\bigg)^{\gamma - 1} + \bigg(\frac{k+1}{k+3}\bigg)^{\gamma - 1} + \ ... \ + \bigg(\frac{k+1}{N}\bigg)^{\gamma - 1} .%\label{secondsum}
\end{eqnarray*}
Observing that if $\gamma > 1$,
\begin{eqnarray*}
\bigg(\frac{k}{k+j}\bigg)^{\gamma - 1} <  \bigg(\frac{k+1}{k+j}\bigg)^{\gamma - 1}  
\end{eqnarray*}
holds true for $j = k+2, ... , N$, 
we may conclude that (\ref{wanted}) certainly holds true provided
\begin{eqnarray}
\bigg(\frac{k}{k+1}\bigg)^{\gamma - 1} <    C_{j^*} - 2 ,
\label{lastwanted}
\end{eqnarray}
and this is true because
$C_{j^*} > 3$ provided $N \ge 5$ (a condition which is satisfied by the networks we are considering) and the left hand side in
(\ref{lastwanted}) is less than $1$.
In conclusion, $k_{nn}(k)$ increases as $k$ increases.

\section*{Acknowledgement}
We would like to thank the anonimous referees for useful remarks which have given us the opportunity to improve the paper.
%, in particular concerning numerical solutions and analytical proofs.


\begin{thebibliography}{100}

\bibitem{Bass}
Bass, F.M., A new product growth for model consumer
durables, Management Sci. 15 (1969) 215-227.

\bibitem{Meade}
Meade, N. and T. Islam,
Modelling and forecasting the diffusion of
innovation -- A 25-year review,
Int. J. of Forecasting 22 (2006) 519-545.

\bibitem{Schi}
Schilling, M.A., ``Strategic management of technological innovation'', McGraw-Hill, 2012.

\bibitem{Rogers}
Rogers, E.M., ``Diffusion of innovations'', Simon and Schuster, 2010.

\bibitem{Christensen}
Bower, J.L. and C.M. Christensen, Disruptive Technologies: Catching the Wave, Harvard Business Review 73 (1995) 43-53.

\bibitem{Phelps}
Phelps, E.S., ``Mass flourishing: How grassroots innovation created jobs, challenge, and change'', Princeton University Press, 2013.

\bibitem{Abr}
Abrahamson, E. and L. Rosenkopf, Social network effects on the extent of innovation diffusion: A computer simulation, Organization Science 8 (1997) 289-309.

\bibitem{Bog1}
Boguna, M., Pastor-Satorras, R. and A. Vespignani, Absence of epidemic threshold in scale-free networks with degree correlations, Phys. Rev. Lett. 90 (2003) 028701.

\bibitem{Bog2}
Barrat, A., Barthelemy, M. and A. Vespignani, ``Dynamical Processes on Complex Networks'', Cambridge Academic Press, 2008.

\bibitem{Boc}
Boccaletti, S., Latora, V., Moreno, Y., Chavez, M. and Hwang, D.-U., Complex networks: Structure and dynamics, Phys. Reports 424 (2006) 175-308.

\bibitem{Wat}
Watts, D.J. and P.S. Dodds, Influentials, networks, and public opinion formation, J. of Consumer Research 34 (2007) 441-458.

\bibitem{Gal}
Di Matteo, T., Aste, T. and M. Gallegati, Innovation flow through social networks: productivity distribution in France and Italy, Eur. Phys. J. B 47 (2005) 459-466.

\bibitem{Ger}
Ghanbarnejad, F., Gerlach, M., Miotto, J.M. and E.G. Altmann,
Extracting information from S-curves of language change,
J. R. Soc. Interface 11 (2014) 20141044.

\bibitem{Gle}
Gleeson, J.P., High-accuracy approximation of binary-state dynamics on networks, Phys. Rev. Lett. 107 (2011) 068701.

\bibitem{Mac}
McCullen, N.J., Rucklidge, A.M., Bale, C.S., Foxon, T.J. and W.F. Gale,  Multiparameter models of innovation diffusion on complex networks, SIAM J. on Applied Dynamical Systems, 12 (2013) 515-532.

\bibitem{Ste}
Rahmandad, H. and J. Sterman, Heterogeneity and network structure in the dynamics of diffusion: Comparing agent-based and differential equation models, Management Science 54 (2008) 998-1014.

\bibitem{VespNaturePh}
Vespignani, A.,
Modelling dynamical process in complex socio-technical systems,
Nature Physics 8 (2012) 32-39. 

\bibitem{Newman}
Newman, M.E.J., Mixing patterns in networks, Phys. Rev. E 67 (2003) 026126.

\bibitem{NewmanSIAM}
Newman, M.E.J., 
The structure and function of complex networks,
SIAM Review 45 (2003) 167-265. 

\bibitem{Gold}
Goldenberg, J., Han, S., Lehmann, D.R. and Hong J.W.,
The role of hubs in the adoption process, Journal of Marketing 73 (2009) 1-13.

\end{thebibliography}
\end{document}